\documentclass{emulateapj}

\def\be{\begin{eqnarray}}
\def\ee{\end{eqnarray}}
\def\etal{et al.}

\def\app{_{\rm app}}
\def\h{_{\rm H}}
\def\hi{_{\rm HI}}
\def\Q{_{\rm Q}}
\def\phs{_{\rm phs}}
\def\phsi{_{{\rm phs},i}}
\def\rec{_{\rm rec}}

\def\Mpc{{\rm\,Mpc}}
\def\yr{{\rm\,yr}}
\def\cm{{\rm\,cm}}
\def\ps{{\rm\,s}^{-1}}

\begin{document}
\title{The apparent shape of the ``Str\"omgren sphere''
around the highest redshift QSOs with Gunn-Peterson troughs }
\author{Qingjuan Yu\footnotemark[1]}
\affil{Department of Astronomy, 601 Campbell Hall, University of California 
at Berkeley, Berkeley, CA 94720; yqj@astro.berkeley.edu}
\footnotetext[1]{Hubble Fellow}

\begin{abstract}

Although the highest redshift QSOs ($z>6.1$) are embedded in a significantly
neutral background universe (mass-averaged neutral hydrogen fraction $>1\%$) as
suggested by the Gunn-Peterson absorption troughs in their spectra, the
intergalactic medium in their vicinity is highly ionized. The highly ionized
region is generally idealized as spherical and called the Str\"omgren sphere.
In this paper, by combining the expected evolution of the Str\"omgren sphere
with the rule that the speed of light is finite, we illustrate the apparent
shape of the ionization fronts around the highest redshift QSOs and its
evolution, which depends on the age, luminosity evolution, and 
environment of the QSO (e.g., the hydrogen reionization history).  The
apparent shape may systematically deviate from a spherical shape, unless the
QSO age is significantly long compared to the hydrogen recombination process
within the ionization front and the QSO luminosity evolution is significantly
slow. Effects of anisotropy of QSO emission are also discussed.  The apparent
shape of the ``Str\"omgren sphere'' may be directly mapped by transmitted
spectra of background sources behind or inside the ionized regions or by
surveys of the hyperfine transition ($21\cm$) line emission of neutral
hydrogen.
\end{abstract}
\maketitle

\section{Introduction}\label{sec:intro}

The existence of the Gunn-Peterson absorption trough in the spectra of the
highest redshift QSOs ($z>6.1$) suggests that hydrogen in the early universe is
significantly neutral ($>1\%$ in mass average; e.g., \citealt{Fanetal02}; see
also \citealt{WL04a}; \citealt{MH04,YL04b}).
However, hydrogen in the
vicinity of the QSOs may be highly ionized by ionizing photons from QSOs (i.e.,
the proximity effect; e.g., \citealt{Bajtlik88}), as indicated in the QSO
spectra by the transmission of flux at wavelengths between the Gunn-Peterson
trough and the Ly$\alpha$ line center. These highly ionized regions around the
QSOs are generally idealized as spherical in the rest frame of the QSO and
called Str\"omgren spheres.\footnote{These highly ionized HII regions (or
Str\"omgren spheres) are well defined only for QSOs that have the Gunn-Peterson
trough. QSOs with relatively low redshifts do not have an unambiguous
Str\"omgren sphere because their environment has already been highly ionized by
stars \citep[e.g., see][]{A04}.} The expansion of the Str\"omgren sphere
depends on the QSO luminosity evolution and its environment \citep[e.g.,][]{SG87,DS87}.  Observers on
Earth perceive an object (here the highly ionized region) by the light reaching
Earth at the same moment.  Since the speed of light is finite, the apparent
shape of the expanding ionization front around the QSO may not be spherical.

\citet{WL04b} show that the ionization fronts along the line of sight and
transverse to it may have different observed sizes (see also discussions
on the size along the line of sight in \citealt{White03} and related
discussions in \citealt{P02}).
In this paper we generalize the time-delay effect
to all angles around the highest redshift QSOs and illustrate the expected
apparent shape of the surrounding ionization front and its evolution
(for the relativistic time-delay effects in other astrophysical contexts,
e.g., see also \citealt{R66,C39}).  We show
how the shape depends on the QSO luminosity evolution and its environment
(such as the neutral hydrogen fraction and clumpiness).
Comparison of observations with the model prediction of the apparent shape
may provide constraints on QSO
properties, the reionization history of the universe, and/or the
cosmological parameter $\Omega_\Lambda$ (by the Alcock-Paczy\'nski
test; \citealt{AP79}).

This paper is organized as follows. In \S~\ref{sec:geom}, we illustrate the
geometry of the time-delay effect on the apparent shape of the ionization front
around a QSO.  In \S~\ref{sec:Rsph}, we study the evolution of the Str\"omgren
sphere in the QSO rest frame.  We combine the evolution and the time-delay
effect to illustrate the evolution of the apparent shape of the ionization
front in \S~\ref{sec:results}.  The effects of anisotropy of QSO emission
are discussed in \S~\ref{sec:aniso}.  Possible observational
tests of the apparent shape are discussed in \S~\ref{sec:test}.  Our
conclusions are summarized in \S~\ref{sec:conc}.

In reality the boundary of an ionized region may not be exactly round even
without considering the relativistic time-delay effect and the QSO anisotropic
emission.  Since the Str\"omgren sphere is idealized as spherical by ignoring
other possible random fluctuations on the boundary, the apparent shapes shown
in this paper are also idealized as smooth and may more appropriately represent the average over the fluctuations.

\section{Geometry}\label{sec:geom}

\begin{figure} \epsscale{1.0} \plotone{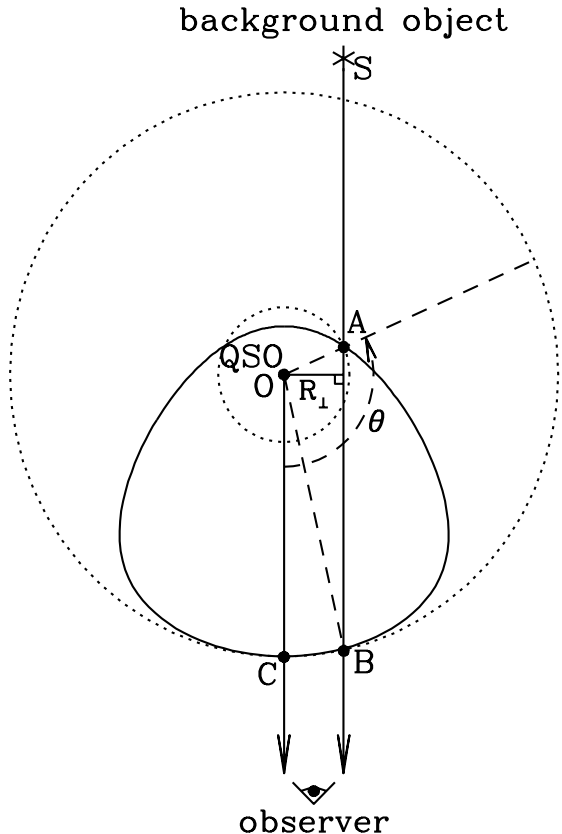} \caption{Schematic
diagram for the apparent shape of the ionization front around a QSO (with       the Gunn-Peterson trough in the spectrum).  The QSO                             is at point $O$ and the distant observer on Earth is located in the
direction of $\vec{OC}$. The QSO emission is assumed to be                      isotropic. The dotted circles represent the ionization fronts at                a given cosmic time ($t_A$ and $t_B$ for inside and outside circles,
respectively) in the rest frame of the QSO.  Hydrogen inside the circle (or the Str\"omgren sphere) is highly ionized by QSO photons. The solid curve is the    apparent ionization front, which follows equation (\ref{eq:geom}).  Photons
emitted from (or passing through) the apparent ionization front (e.g., from     point $A$ at $t_A$ and $B$ at $t_B$ shown in this figure) reach the             observer at the same time. The apparent shape of the ionization front is
rotationally symmetric along $OC$ and is generally not a                        sphere/spheroid/paraboloid. The apparent shape could be mapped by observational spectra of background objects (e.g., $S$ in the figure) or
the tomograph of HI $21\cm$ line emission.                                      }                                                                               \label{fig:f1}
\end{figure}

We use Figure~\ref{fig:f1} to illustrate the geometry of the apparent shape of
the ionization front around a QSO at point $O$. For simplicity, we ignore the
Hubble expansion within the ionization front, since here the physical size of
the ionized region around the highest redshift QSOs is much smaller than the
Hubble scale and the characteristic timescale of the QSO luminosity
evolution is much smaller than the age of the universe at $z\sim 6$.
The derivation below is a simple
application of the rule that the speed of light is finite.  Suppose that the
nuclear activity of the QSO is triggered at cosmic time $t_i$. The ionizing
photons emitted from the QSO will then ionize surrounding neutral hydrogen.
For simplicity, we assume that the QSO emission is isotropic here and discuss
the anisotropic effect of the emission in \S~\ref{sec:aniso}. In
the rest frame of the QSO, a spherical ionization front will form and expand
(see the dotted circles in Fig.~\ref{fig:f1}), separating the inside highly
ionized region and the outside neutral or partly neutral region. We denote the
physical (proper) radius of the expanding sphere at cosmic time $t$ ($\ge t_i$) by
$R(t;t_i)$. The detailed evolution of $R(t)$ is related to the QSO luminosity
evolution and its environment, which 
will be determined in \S~\ref{sec:Rsph}.  The observer on Earth
perceives the ionization front by the light reaching Earth at the same
moment.  If photons emitted from (or passing through) the ionization front, e.g.,
point $A$ in Figure~\ref{fig:f1}, reach the distant observer at the same time
as photons from the QSO, the photons from $A$ should be emitted from (or pass
through) $A$ at cosmic time 
\be
t_A=t(z_Q)+R(t_A)\cos\theta/c, \qquad 0\le\theta\le\pi,
\label{eq:tA}
\ee
where
$z_Q$ is the QSO redshift, $t(z_Q)$ is the cosmic time corresponding to $z_Q$,
$c$ is the speed of light, and $\theta$ is the angle between $\overrightarrow{OA}$ and the
observer's line of sight $\overrightarrow{OC}$ (see Fig.~\ref{fig:f1}).  For QSO photons
to reach the ionization front $A$ at cosmic time $t_A$, they should emit from
the QSO $O$ at an earlier time
\be
t_O=t_A-R(t_A)/c.
\label{eq:tO}
\ee
Given any $t_O$ ($t_i\le t_O\le t_A$), the solution
of equation (\ref{eq:tO}) $t_A$ is unique since the expansion speed of
$R(t_A)$ is slower than $c$.
Thus we may define such a function $r(\tau;t_i)\equiv R(t_A;t_i)$ by changing
the independent variable from $t_A$ to $\tau$, where $\tau\equiv t_O-t_i$ is
the age of the QSO at $t_O$.
According to the above, the apparent shape of the
ionization front ($r,\theta$) should satisfy the following equation:
\be
r(\tau)(1-\cos\theta)=c(\tau_Q-\tau), \qquad 0\le\theta\le\pi,
\label{eq:geom} \ee
where $\tau_Q\equiv t(z_Q)-t_i$ is the age of the QSO at cosmic time $t(z_Q)$.
The apparent shape (illustrated by the solid
curve in Fig.~\ref{fig:f1}) is rotationally symmetric along the observer's
line of sight to the QSO.  The shape is generally not a sphere
[unless $r(\tau)$ does not change with $\tau$].
According to equation (\ref{eq:geom}), given $\tau_Q$ and an increasing
function $r(\tau)$, the apparent size $r$ decreases with increasing $\theta$.
Because of the existence of the term ``$-\tau$'' in the right-hand side of equation
(\ref{eq:geom}), the apparent shape $(r,\theta)$ is neither a spheroid (the
apparent shape of relativistically expanding spherical shell of radio sources
discussed in \citealt{R66,R67}) unless $r(\tau)$ is a linear function of $\tau$
nor a ``light-echo'' paraboloid around a nova or supernova \citep{C39}.

We define the apparent radial expansion speed of the ionization front at 
a given $\theta$ by $v\app\equiv dr'/dt'$, where $r'=(1+z)r$ is 
the comoving distance of $r$ and $dt'=(1+z)d\tau_Q$ is the time interval measured 
by the observer with including the effect of the cosmological time dilation.
According to equations (\ref{eq:tA})--(\ref{eq:geom}), the apparent radial
expansion speed
of the ionization front in units of $c$ at a given $\theta$ is given by
\begin{eqnarray}
\beta\app\equiv \frac{v\app}{c}=\left.\frac{1}{c}\frac{dr}{d\tau_Q}\right|_\theta & = & \frac{dr/d\tau}{c+(dr/d\tau)(1-\cos\theta)}, \label{eq:betaapp}\\
& = & \frac{dR/dt_A}{c-(dR/dt_A)\cos\theta}.
\label{eq:speed}
\end{eqnarray}
Note that the physical expansion speed of the ionization front in the QSO rest
frame is 
\be
dR/dt_A=(dr/d\tau)/(1+c^{-1}dr/d\tau)<c,
\label{eq:dRdtA}
\ee
which is consistent with
special relativity that real information does not travel faster than the speed
of light. If $dR/dt_A$ is highly relativistic,
the apparent expansion is superluminal at sufficiently small $\theta$, i.e.,
$\beta\app>1$, and $\beta\app$ is close to $1/2$ when $\theta\rightarrow\pi$.
We can also define the apparent transverse expansion speed in units of
$c$ through $\beta\app{_{,\perp}}\equiv\beta\app \sin\theta$, which gives
the conventionally transverse superluminal phenomenon at sufficiently small
$\theta$ if $dR/dt_A\rightarrow c$.
According to equation (\ref{eq:speed}), the lower limit on the QSO lifetime can
be constrained simply by the apparent sizes of the ionization front, e.g.,
$\tau_Q>2r(\theta=\pi)/c$ or $\tau_Q>r(\theta=\pi/2)/c$.

\section{The evolution of $\lowercase{r}(\tau)$} \label{sec:Rsph}

We denote the mean neutral
hydrogen fraction in the region surrounding the QSO prior to the triggering of
its nuclear activity by $x\hi\equiv\langle
n\hi\rangle/\langle n\h\rangle$ ($\sim$ several percent to 10--20\% in
\citealt{YL04b}; and the rest of the hydrogen has been ionized by stars or collisionally ionized in large halos; also cf., \citealt{MH04} or for a higher
fraction \citealt{WL04a}), where $\langle
n\hi\rangle$ is the average proper number density of neutral hydrogen and
$\langle n\h\rangle$ is the density of total hydrogen.  After the nuclear
activity is triggered, the expansion of the ionization front (or the
Str\"omgren sphere) in the rest frame of the QSO as a function of $\tau$
(not as a function of $t_A$ here) may be described by the following equation
(e.g., \citealt{SG87,DS87}; \citealt{CH00,MR00,YL04b}):
\be
\frac{4\pi}{3}\frac{d(x\hi\langle n\h\rangle r^3)}{d\tau}=
\dot{N}\phs(\tau) 
-\frac{4\pi}{3}\alpha_{\rm B}C \langle n\h\rangle^2 r^3,
\label{eq:stromevol}
\ee
where $\dot{N}\phs(\tau)$ is the ionizing photon emission rate of the QSO
at its age
$\tau$, $\alpha_{\rm B}$($=2.6\times 10^{-13}\cm^3\ps$ at $T=10^4$K) is the
hydrogen recombination coefficient to excited levels of hydrogen, $C\equiv
\langle n\hi^2\rangle/\langle n\h\rangle^2$ is the clumping factor
describing the effective clumpiness of hydrogen ionized by QSO photons,
and ``$\langle\cdot\cdot\cdot\rangle$'' represents the space average over the
sphere with radius $r$.
Note that here $C$ is a little different from the conventional definition of
$\langle n\h^2\rangle/\langle n\h\rangle^2$.
The second term on the right-hand side of equation
(\ref{eq:stromevol}) accounts for 
the recombination of ionized hydrogen within the Str\"omgren sphere.
In equation (\ref{eq:stromevol}) we set the parameters
($x\hi,\langle n\h\rangle,C$) to be the values at cosmic
time $t=t_i+\tau$,
and their changes during the time interval $r/c$ of the QSO photon propagation
in the Str\"omgren sphere have been assumed to be negligible.
For simplicity, we also neglect their
dependence on $r$ and set them to be constants in the calculations below.
Thus, the solution of equation (\ref{eq:stromevol}) is given by
\be
r(\tau)=\left[\frac{3}{4\pi\langle n_H\rangle x\hi}
\int_0^\tau \dot N\phs(\tau')\exp\left(\frac{\tau'-\tau}{\tau\rec}\right)d\tau'
\right]^{1/3},
\label{eq:rsol}
\ee
where the recombination timescale
\be
\tau\rec&\equiv&x\hi(C \langle n\h\rangle \alpha_{\rm B})^{-1} \nonumber\\
&\simeq&4\times 10^6\yr\left(\frac{x\hi}{0.1}\right)\left(
\frac{30}{C} \right) \left(\frac{7.4}{1+z} \right )^3
\label{eq:trec}
\ee
characterizes the role of the recombination term in equation
(\ref{eq:stromevol}). 

The physical expansion of the ionization front in the rest frame of the QSO
$dR/dt_A$ was discussed in \citet{White03} by accounting for finite light
travel time and ignoring the recombination terms, and was generalized in equation
(1) of \citet{WL04a}.
Here applying equation (\ref{eq:stromevol}) into equation (\ref{eq:dRdtA}),
we obtain the physical expansion of the ionization front as follows:
\be
\frac{dR}{dt_A}=c\frac{\dot N\phs-\frac{4\pi}{3}\alpha_{\rm B}C \langle n\h\rangle^2 R^3}{4\pi c x\hi\langle n\h\rangle R^2+\dot N\phs-\frac{4\pi}{3}\alpha_{\rm B}C \langle n\h\rangle^2 R^3},
\ee
where $\dot N\phs=\dot N\phs(\tau)=\dot N\phs(t_A-t_i-R/c)$.
At $\theta=0$ we have $\tau=\tau_Q$ (eq.~\ref{eq:geom})
and thus $\dot N\phs$ is simply $\dot N\phs(\tau_Q)$;
at $\theta=\pi/2$ we have $\tau=\tau_Q-R/c$ and thus $\dot N\phs$ is
simply $\dot N\phs(\tau_Q-R/c)$.
Note that in principle the apparent ionization front at $\theta=\pi/2$ may not
necessarily correspond to the maximum angle of the apparent ionized region
extended on the sky (see Fig.~\ref{fig:f1}).

Note that in equation (\ref{eq:stromevol})
$\dot{N}\phs(\tau)$ is a non-decreasing function of $\tau$ so that $r(\tau)$
expands with increasing $\tau$.  When $\dot{N}\phs(\tau)$ decreases with
increasing $\tau$ (e.g., close to the quenching of the QSO nuclear activity)
and $\tau\gg\tau\rec$,
the ionization front shrinks, and its evolution may not be described by
equation (\ref{eq:stromevol}) if the characteristic decreasing timescale of
$\dot{N}\phs(\tau)$ is significantly short compared to the recombination
timescale.  For simplicity, in this paper we ignore the decreasing phase of the
QSO luminosity evolution, which has been shown to at least not dominate the
main population of luminous QSOs observed at relatively low redshifts
\citep{YL04a}.

Below we assume two simple models of $\dot{N}\phs(\tau)$
\citep[see also][]{YL04b}: in model (i) $\dot{N}\phs(\tau)=\dot{N}\phsi$ is a
constant, and in model (ii) $\dot{N}\phs(\tau)=\dot{N}\phsi\exp(\tau/\tau_S)$
increases exponentially with $\tau$, where $\tau_S\simeq 4.5\times
10^7\yr[0.1\epsilon/(1-\epsilon)]$ is the Salpeter timescale and $\epsilon$ is
the mass-to-energy conversion efficiency. 
When $\tau\ll\tau\rec$ in model (i) or $\tau\ll\min\{\tau\rec,\tau_S\}$ in
model (ii),
the effects of recombination and the evolution of the QSO luminosity are
both negligible and $r(\tau)$ is given by
\be
r(\tau)\simeq\left( \frac{3\dot{N}\phsi\tau}
{4\pi x\hi\langle n\h\rangle} \right)^{1/3}=r_S\left(\frac{\tau}{\tau\rec}\right)^{1/3},
\label{eq:strom1}
\ee
where
\be
r_S\equiv \left(\frac{3\dot{N}\phsi\tau\rec}{4\pi x\hi\langle n\h\rangle}\right)^{1/3}=\left( \frac{3\dot{N}\phsi}{4\pi\alpha_{\rm B}
C\langle n\h\rangle^2}\right)^{1/3}.
\label{eq:rS}
\ee
Given equation (\ref{eq:strom1}), the changing rate of $r(\tau)$ with
$\tau$ is larger than  $c$ if $\tau\la [r_S/(3c\tau\rec)]^{3/2}\tau\rec$
(i.e., $r(\tau)\ga 3c\tau$ here).
As the ionization front expands, if $\tau\gg\tau\rec$ (and
$\tau\rec\ll\tau_S$ in model ii), the recombination
within the highly ionized HII region is approximately balanced by the emission
of the ionizing photons from the QSO, and we have
\be
r(\tau)\simeq \cases{r_S & for model (i), \label{eq:strom2} \cr
r_S\left(\frac{\tau_S}{\tau\rec+\tau_S}e^{\tau/\tau_S}\right)^{1/3}
& for model (ii), \label{eq:strom2s} }.
\ee
which is insensitive to the detailed value of $\tau$ for model (i) and depends
on $\tau$ only through $N\phs(\tau)$ in model (ii).  
In case that $\tau\rec\gg\tau_S$ in model (ii), $r(\tau)$ at $\tau\gg\tau_S$
still follows equation (\ref{eq:strom2s}), but the effect of recombination is
not important and $r(\tau)$ is mainly determined by the QSO emission. Given 
equation (\ref{eq:strom2s}), for the exponential increase of the luminosity
evolution in model (ii), the changing rate of $r(\tau)$ with $\tau$ is larger
than $c$ if $r(\tau)\ga 3c\tau_S$.

\section{The evolution of the apparent shape of the ionization front}
\label{sec:results}

\begin{figure} \epsscale{1.0} \plotone{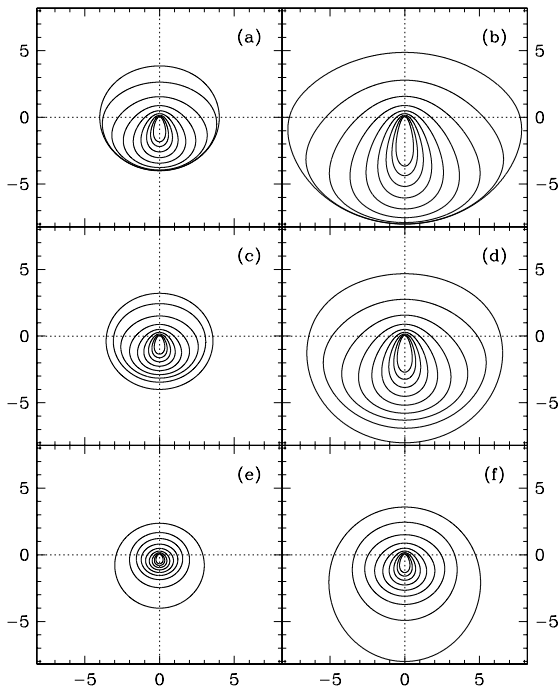} \caption{Expected evolution
of the apparent shape of the ionization front around a QSO. The geometric
configuration of the QSO (located at the crossing point of the dotted lines)
and the observer is the same as in Figure~\ref{fig:f1}.  The QSO emission is
assumed to be isotropic. Both of the axes are in units of $c\tau\rec$ (see
$\tau\rec$ in eq.~\ref{eq:trec}).  Top panels (a)-(b) are for model (i) of
$\dot{N}\phs$ (see \S~\ref{sec:results}), and middle and bottom panels for
model (ii) with $\tau_S/\tau\rec=10$ (c-d) and 3 (e-f), respectively.  The
parameters ($\dot{N}\phsi,\langle n\h\rangle,C$) are set so that
$r(\tau=10\tau\rec)/(c\tau\rec)=4$ in the left panels (a,c,e) and 8 in the right panels
(b,d,f). In each panel, different curves represent the apparent shapes with
different $\tau_Q/\tau\rec$. The $\tau_Q/\tau\rec$ increases from inward to
outward with $\delta\log(\tau_Q/\tau\rec)=0.25$, and the outmost curve has
$\tau_Q/\tau\rec=10$. } \label{fig:model1} \end{figure}

By combining equations (\ref{eq:geom}) and (\ref{eq:rsol}), we obtain the
evolution of the apparent shape of the ionization front ($r,\theta$) for
different models of the QSO luminosity evolution, which is shown in
Figure~\ref{fig:model1}. The evolution of the apparent shape and size
depends on the QSO luminosity evolution and its environment
(e.g., the hydrogen reionization history or structure formation)
through the parameters ($\tau_Q,\tau_S,\tau\rec,r_S$).  
In Figure~\ref{fig:model1}, the parameters ($\dot{N}\phsi,x\hi,C,\langle
n\h\rangle$) are set so that $r_S/(c\tau\rec)$ is larger in the right
panels than in the left ones. 
Top panels (a) and (b) are for model (i) of
$\dot{N}\phs(\tau)$. As seen from them, the apparent shapes of the ionization
fronts are generally not a sphere, especially if $\tau_Q$ is
short (inner curves) and $r_S/(c\tau\rec)$ is large (e.g.,
because $\dot{N}\phsi$ is
high or $x\hi$ is low; right panel).  The apparent shape expands
as $\tau_Q$ increases. When $\tau_Q$ is long enough, the apparent shape at
small $\theta$ (see the definition of $\theta$ in Fig.~\ref{fig:f1}) starts to
be almost independent of $\tau_Q$ and approaches a sphere, since the radius
of the Str\"omgren sphere approaches the equilibrium value $r_S$
(eq.~\ref{eq:strom2}) at which the ionizing photon emission is balanced by the
recombination.  With increasing $\tau_Q$, the shape at large $\theta$ continues
to expand and finally the whole apparent shape becomes spherical (see the
outmost curve in panel a). According to equation (\ref{eq:rS}), the stronger
the $\dot{N}\phsi$ or the smaller the $C$, the larger the final equilibrium
sphere radius. In addition, in case that $r_S/(c\tau\rec)$ is significantly
small ($\ll 1$), the expansion may become non-relativistic at a time much
earlier than $\tau_Q\simeq \tau\rec$, thus the apparent shape may
approach a sphere even if $\tau_Q\ll\tau\rec$ (but $\tau_Q>r(\tau_Q)/(3c)$).
Figure~\ref{fig:model1}c-f illustrate the apparent shapes for
model (ii) of $\dot{N}\phs(\tau)$, where the timescales of increasing
luminosity are set to have $\tau_S/\tau\rec=10$ and 3, respectively.  
In model (ii) the size of the apparent shape is sensitive
to $\tau_Q$ even if $\tau_Q$ is long because $\dot{N}\phs$ increases
with time exponentially; and the shape does not evolve towards a spherical
shape, especially if $\tau_S$ is small (bottom panels e-f). 

For the realistic case of the highest redshift QSOs with Gunn-Peterson troughs
observed so far ($z\sim 6.1-6.4$; \citealt{White03,Fan04}), \citet{YL04b} study
the environmental effect of their surrounding regions and obtain $x\hi\sim 0.1$
and $C\sim 30$--40 (which are shown to be consistent with the observed
Str\"omgren radii). Then the recombination timescale $\tau\rec$ is $\sim 4\times
10^6\yr$ (eq.~\ref{eq:trec}), which is significantly shorter than the QSO
lifetime ($\ga 4\times 10^7\yr$; e.g., \citealt{YL04a}).  Thus the ages of most
observed QSOs should be long enough that the sizes of the ionization
fronts along the line of sight are close to the equilibrium values. For the
observed QSOs with
$\dot{N}\phs(\tau_Q)\sim 10^{57}\ps$, the apparent size at $\theta=0$ is
$r(\tau\Q)\sim 5\Mpc$($\sim3-4c\tau\rec$) \citep{YL04b}, and the apparent
shapes should be close to the outer curves shown in Figure~\ref{fig:model1}a
and c (see also Fig.~\ref{fig:aniso} if the QSO emission is anisotropic).
Statistically a small fraction of observed QSOs would also have
significantly short ages and the apparent shapes of their ionization fronts
should systematically deviate from a spherical shape (see the inner curves in
Fig.~\ref{fig:model1}).
\citet{WL04a} obtain a much higher $x\hi$ and a low QSO age from the
observed Str\"omgren radius
of a QSO by using a much smaller clumping factor; in this case, the
recombination timescale is significantly long (up to $10^9\yr$) and $r(\tau_Q)$
is not significantly smaller than $3c\tau_Q$, so the apparent shapes of
most observed QSOs should be more likely to systematically deviate from a
spherical shape.

\begin{figure} \epsscale{1.0} \plotone{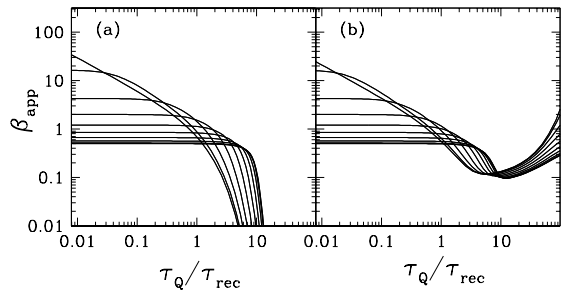} \caption{Expected
apparent radial expansion speed of the ionization front in units of $c$
as a function of $\tau_Q/\tau\rec$ (eq.~\ref{eq:betaapp}).
Panel (a) (model i of $\dot{N}\phs$) has the same parameters as those in
Figure~\ref{fig:f1}a, and panel (b) (model ii) the same as those in
Figure~\ref{fig:f1}c.  In each panel, different curves give the
speeds at different $\theta$, which increases from 0 to $180^\circ$ with
$\delta\theta=30^\circ$ from top to bottom at the low-$\tau_Q/\tau\rec$
end.  The apparent speed at small $\theta$ is superluminal at small
$\tau_Q/\tau\rec$ in both panels and also at large $\tau_Q/\tau\rec$ for model
(ii) in panel (b). For simplicity, we do not show the apparent transverse
expansion speed $\beta_{\rm app,\perp}=\beta\app\sin\theta$ here, which
also gives the similar superluminal phenomenon except at $\theta=0$.
} \label{fig:speed} \end{figure}

Using equation (\ref{eq:speed}), we illustrate the apparent expansion speed of
the ionization front as a function of $\tau_Q$ in Figure~\ref{fig:speed}.
Panel (a) is for model (i) of $\dot{N}\phs$ and panel (b) for model (ii).  As
seen from Figure~\ref{fig:speed}, the speed is superluminal at small $\theta$
and $\tau_Q/\tau\rec$.  For the specific case of $\theta=0$, we have
$\tau=\tau_Q$ and $\beta\app=c^{-1}dr/d\tau$ (eqs.~\ref{eq:geom} and
\ref{eq:betaapp}), and the corresponding superluminal expansion is also
mentioned or discussed in some recent literature \citep{CH00,White03,WL04b}.
In panel (b), the speed at small $\theta$ may also become superluminal even at
large $\tau_Q/\tau\rec$ because the ionizing photon emission rate and
$r(\tau)$ (eq.~\ref{eq:strom2s}) increases exponentially in model (ii).  The
expansion of the highly ionized regions around the highest redshift QSOs with
Gunn-Peterson troughs provides another astrophysical example of the
superluminal phenomenon (for other examples, e.g., see reviews in
\citealt{BMR77,FFP03}).

\section{Effects of anisotropy of QSO emission}\label{sec:aniso}

In the above, QSO ionizing photons have been assumed to be emitted
isotropically. However, in reality the QSO emission may be anisotropic at least
in these two cases: (i) a torus is surrounding the central black hole in the
QSO, which blocks the ionizing photons from the central engine; (ii) the UV
ionizing photon emission from the accretion disk surrounding the black hole is
anisotropic (see also discussions of the effects of anisotropy of QSO emission
in \citealt{Cen03} and \citealt{WL04b}). 

In case (i), the highly ionized region in the rest frame of the QSO may appear
simply like a large-scale version of the [OIII] ionization ``cone'' detected in
local Seyfert 2 galaxies. The expected apparent shape is easily correspondingly
adjusted depending on the normal direction and the opening angle of the torus.

\begin{figure} \epsscale{1.0} \plotone{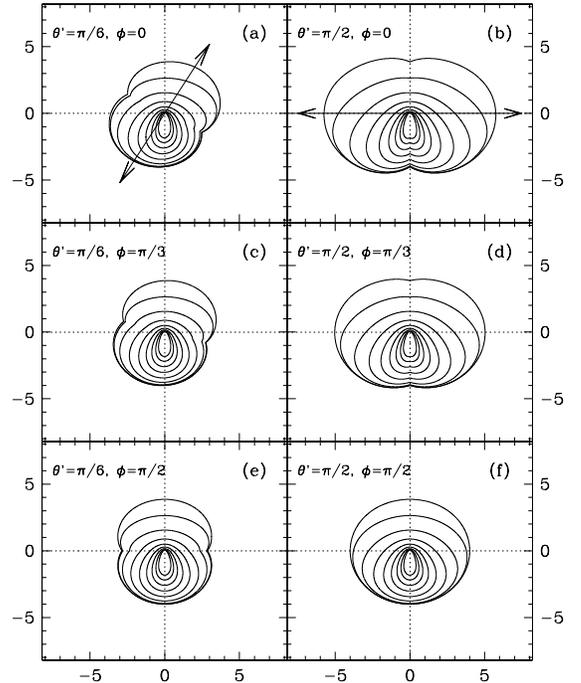} \caption{
Cross sections of the apparent ionized region after including the anisotropy of
the accretion disk emission due to the limb-darkening effect in
Figure~\ref{fig:model1}a.  The solid lines with arrows in (a) and (b) are the
disk normal passing the QSO. The emission is assumed to be symmetric about the
disk plane and rotationally symmetric about the disk normal.  The parameters
($\dot{N}\phsi,\langle n\h\rangle,C$) are set so that
$r[\tau(10\tau\rec,\theta=0]=4$.  After including the anisotropic effect, the
rotational symmetry about the observer's line of sight mentioned in
Figure~\ref{fig:f1} is broken, so the cross sections are different in the top,
middle, and bottom panels.  As in Figure~\ref{fig:f1}, for each curve, the
apparent size in any direction from the central QSO with $\theta<\pi/2$ is not
smaller than the size in the opposite direction due to the time-delay effect.
See details in \S~\ref{sec:aniso}.  } \label{fig:aniso} \end{figure}

In case (ii), we assume that the emergent intensity from the accretion disk is
$\propto (1+2|\mu|)$ because of the limb-darkening effect (e.g., \citealt{K99}),
where $\mu$ is the cosine of the angle between the disk normal and the emitting
direction (so far there is no much evidence to show that UV photons from QSOs
are strongly beamed). Here the QSO anisotropic emission is symmetric about the
disk plane and axisymmetric about the disk normal.  Although both the
relativistic time-delay and the anisotropy of the QSO emission may make the
apparent shape deviate from a spherical shape, their effects are different.  As
mentioned above, the time-delay effect decreases $r(\tau)$ with increasing
$\theta$ (see eq.~\ref{eq:geom}) and does not break the rotational symmetry of
the ionized region about the observer's line of sight $\overrightarrow{OC}$.
For the anisotropic effect here, the detailed dependence of the apparent size
on $\theta$ is significantly affected by the detailed geometric configuration
between the accretion disk and the observer. The rotational symmetry of the
apparent shape around the observer's line of sight is generally broken unless
the disk normal is coincident with the observer's line of sight. We denote
the angle between the disk normal and the observer's line of sight by $\theta'$
($0\le\theta'\le\pi/2$). For $\theta'\ne 0$, the cross sections of the apparent
ionized region are illustrated in Figure~\ref{fig:aniso}, where the observer's
line of sight is on the cross section, $\phi$ ($0\le\phi<\pi$) is the angle of
the cross section anti-clockwise from the plane in which $\overrightarrow{OC}$ and
the disk normal are located, and
$|\mu|=|\sin\theta\cos\phi\sin\theta'\pm\cos\theta\cos\theta'|$.  Although in
general $r(\tau)$ does not necessarily decrease with increasing $\theta$ if the
QSO emission is anisotropic, the apparent size in any direction from the
central QSO with $\theta<\pi/2$ is still systematically not larger than the
size in the opposite direction because of the time-delay effect and the
symmetry assumed for the anisotropic emission here.  Given the detailed
discussions in \S~\ref{sec:results}, if the QSO age is significantly long
compared to the hydrogen recombination process and the QSO luminosity evolution
is significantly slow, the apparent ionized region follows the same symmetry as
the QSO emission (i.e., symmetric about the disk plane and axisymmetric about
the disk normal here).  As seen from Figure~\ref{fig:aniso}, the smaller the
$\tau_Q/\tau\rec$ and the smaller the $\theta'$, the less important the
anisotropic effect.

\section{Possible observational tests}\label{sec:test}

We use Figure~\ref{fig:f1} to illustrate the possible observational tests
of the apparent shape. The discussion below may be generalized
to the anisotropic case of QSO emission.

\begin{figure} \epsscale{1.0} \plotone{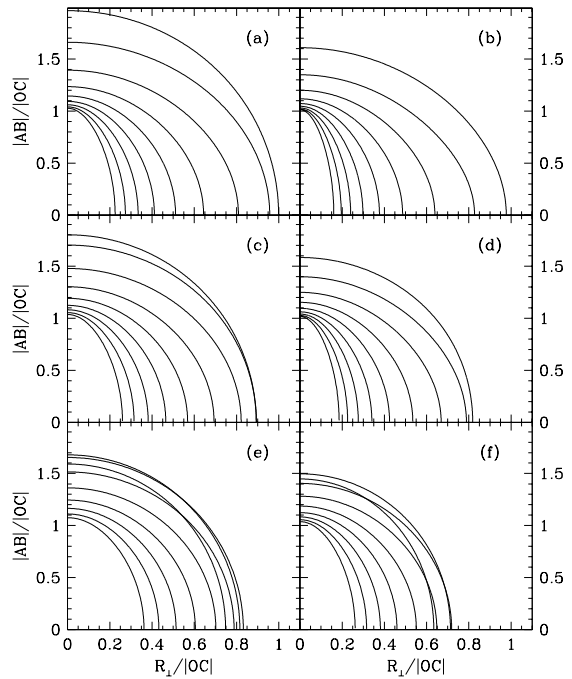} \caption{Expected
ratio of $|AB|/|OC|$ of the apparent ionization front as a function of
$R_\perp/|OC|$ (see Fig.~\ref{fig:f1} for the definition of the proper
distances of $|AB|$, $|OC|$, and $R_\perp$). The parameters are the same as
those in Figure~\ref{fig:model1}.
This figure illustrates that the evolution of the apparent ionization
fronts around QSOs may be inferred from the distribution of $|AB|$.
For observational methods of determining $|AB|$ and $|OC|$, see discussion
in \S~\ref{sec:test}. A similar study may also be done for $|OA|$ and $|OB|$. }
\label{fig:ratio} \end{figure}

If there exists a background object $S$ behind the highly ionized region (see
Fig.~\ref{fig:f1}), significant transmission of Ly$\alpha$ (and/or Ly$\beta$)
photons may be present on the Gunn-Peterson trough in the observational
spectrum of $S$ at wavelengths corresponding to redshifts $z(t_A)$ to $z(t_B)$
[somewhat like (but it may be much broader than) the observed emission spike on
top of the Gunn-Peterson trough of SDSS J1148+5251; see Figs.~4 or 5 in
\citealt{White03}].  Similarly, the spectra of background objects lying inside
the highly ionized regions (see also discussions for inside objects in \citealt{Cen03}) may also show transmission features of Ly$\alpha$
photons at the wavelength corresponding to the redshift when its light travels
out of the ionized region.  If the background object is the QSO itself, we have
$\tau=\tau_Q$ at $\theta=0$ (eq.~\ref{eq:geom}) and the size of the
apparent ionization front $|OC|=r(\tau_Q)$, which can be measured directly from
the wavelength difference between the Gunn-Peterson trough and the Ly$\alpha$
line center in the QSO spectrum \citep[e.g.,][]{White03}.  Note that the
existence of Ly$\alpha$ emitting galaxies lying behind or inside the highly
ionized regions of the highest redshift QSOs is likely and detection of them is
possible by current techniques, since the apparent ``Str\"omgren sphere'' with
a physical size $\sim 5\Mpc$ at $z\sim 6$ covers an area about $14\times14~{\rm
arcmin}^2$ on the sky and is large enough for this compared to the survey areas of some
Ly$\alpha$ surveys (see Table 1 in \citealt{Sternetal}) that have successfully
discovered galaxies at $z\sim 6.5$.  By obtaining observational spectra of a
number of background sources with different projecting distances
($R_\perp=|OA|\sin\theta$) to the highest redshift QSO(s), one can measure the
proper distance of $|AB|$ (or $|OA|$, $|OB|$) for each source behind the region(s) (or $|OB|$
for inside sources) and then statistically map the three-dimensional apparent
shape of the highly ionized region(s) surrounding the QSO(s). The distribution
of these distances as a function of $R_{\perp}$ may be used to statistically
constrain the evolution of the apparent ionization fronts (and hence the QSO
luminosity evolution and its environment), as illustrated in
Figure~\ref{fig:ratio}.

In addition, the apparent shape of the highly ionized region surrounding the
QSOs with Gunn-Peterson troughs might also be revealed by radio surveys of the
hyperfine transition $21\cm$ line emission of neutral hydrogen (e.g., the Low
Frequency Array, the Square Kilometer Array, the PrimevAl Structure Telescope,
the Mileura Wide-Field Array), a possible technique to map the neutral hydrogen
distribution at high redshift \citep[see discussions in][]{WL04b}.

For the observer on Earth, the physical distance of $R_\perp$ can be
obtained by $R_\perp=D_A\delta\phi$, where $D_A$ is the angular diameter
distance to the QSO and $\delta\phi$ is the angle of the QSO and the ionization
front (e.g., point $A$) separated on the sky.  The transformation of the
angular and the redshift space to real space ($R_\perp$ and $|AB|$, $|OA|$ or
$|OB|$ etc.) depends on the cosmological parameters. If the QSO age and
environment are effectively constrained by the apparent ionization front
(through $|OC|$, $|AB|/|OC|$ etc.) at small $R_\perp\sim 0$ and if the
anisotropy of QSO emission is effectively constrained by some cross sections of
the apparent ionized region, the apparent physical sizes of the ionized region
at other directions may be predicted by using the method in this paper.  Thus,
the ratio of the redshift difference between the QSO and the apparent ionization
front at $R_\perp\sim 0$ to the maximum angle(s) of the apparent ionized region
(or cross sections) extended on the sky might be used to provide
constraints on the cosmological parameter $\Omega_\Lambda$ by the
Alcock-Paczy\'nski test \citep{AP79}, since the angular size significantly
depends on $\Omega_\Lambda$ but the redshift size does not at $z\sim 6$.

\section{Conclusions}\label{sec:conc}

In this paper we have illustrated the apparent shape of the
ionization front around the highest redshift QSOs, and its evolution, with Gunn-Peterson troughs in
the spectra.  We have shown that the apparent shape and its evolution depend
on the age and the luminosity evolution of the QSO, the recombination process
of ionized hydrogen surrounding the QSO, and the anisotropic property of
the QSO emission.  Both the relativistic time-delay effect and the anisotropy
of QSO emission may make the apparent shape deviate from a sphere.  The
time-delay effect systematically makes the apparent size of the ionization
front in the directions away from the observer smaller than the size in the
directions towards the observer.  The deviation of the shape caused by the
time-delay effect is not significant if the QSO age is significantly long
compared to the hydrogen recombination process within the ionization front and
the QSO luminosity evolution is significantly slow.  The time-delay effect does
not break the rotational symmetry of the ionized region about the observer's
line of sight to the QSO.  The apparent expansion of the ionization front may
be superluminal at the early stage of its expansion or if the QSO ionizing
photon emission rate increases dramatically (e.g., exponentially).  The
anisotropic effect on the apparent shape depends on the detailed anisotropy of
the emission (e.g., affected by the normal direction of the accretion disk in
the QSO) and may break the rotational symmetry about the observer's line of
sight.

The apparent shape of the highly ionized regions may be mapped
by transmitted spectra of background sources behind or inside them or by surveys
of the hyper-fine transition $21\cm$ line of neutral hydrogen. 
The apparent shape could also be used to constrain the cosmological parameter
$\Omega_\Lambda$ by the Alcock-Paczy\'nski test.

I am grateful to Youjun Lu for helpful discussions and thoughtful comments. 
I thank Ue-Li Pen for helpful discussions.
QY acknowledges support provided by NASA through Hubble Fellowship grant
\#HF-01169.01-A awarded by the Space Telescope Science Institute, which is
operated by the Association of Universities for Research in Astronomy, Inc.,
for NASA, under contract NAS 5-26555.

\end{document}